\documentclass[superscriptaddress,amsmath,amssymb,aps,prl,twocolumn]{revtex4-1}

\usepackage{times}
\usepackage{graphicx}
\usepackage{dcolumn}

\begin{document}

\title{Multiwavelength polarization insensitive lenses based on dielectric metasurfaces with meta-molecules}


\author{Ehsan Arbabi}
\affiliation{T. J. Watson Laboratory of Applied Physics, California Institute of Technology, 1200 E. California Blvd., Pasadena, CA 91125, USA}
\author{Amir Arbabi}
\affiliation{T. J. Watson Laboratory of Applied Physics, California Institute of Technology, 1200 E. California Blvd., Pasadena, CA 91125, USA}
\author{Seyedeh Mahsa Kamali}
\affiliation{T. J. Watson Laboratory of Applied Physics, California Institute of Technology, 1200 E. California Blvd., Pasadena, CA 91125, USA}
\author{Yu Horie}
\affiliation{T. J. Watson Laboratory of Applied Physics, California Institute of Technology, 1200 E. California Blvd., Pasadena, CA 91125, USA}
\author{Andrei Faraon}
\email{Corresponding author: A.F.: faraon@caltech.edu}
\affiliation{T. J. Watson Laboratory of Applied Physics, California Institute of Technology, 1200 E. California Blvd., Pasadena, CA 91125, USA}

\begin{abstract}
Metasurfaces are nano-structured devices composed of arrays of subwavelength scatterers (or meta-atoms) that manipulate the wavefront, polarization, or intensity of light. Like other diffractive optical devices, metasurfaces suffer from significant chromatic aberrations that limit their bandwidth. Here, we present a method for designing multiwavelength metasurfaces using unit cells with multiple meta-atoms, or meta-molecules. Transmissive lenses with efficiencies as high as 72$\%$ and numerical apertures as high as 0.46 simultaneously operating at 915 nm and 1550 nm are demonstrated. With proper scaling, these devices can be used in applications where operation at distinct known wavelengths is required, like various fluorescence microscopy techniques.
\end{abstract}

\maketitle

Over the last few years, a new wave of interest has risen in nano-structured diffractive optical elements due to advances in nano-fabrication technology~\cite{Astilean1998OptLett,Lalanne1999JOSAA,Fattal2010NatPhoton,Lin2014Science,Kildishev2013Science,Yu2014NatMater,Vo2014IEEEPhotonTechLett}. From the multiple designs proposed so far, dielectric transmitarrays~\cite{Vo2014IEEEPhotonTechLett,Arbabi2015NatCommun,Arbabi2015NatNano} are some of the most versatile metasurfaces because they provide high transmission and subwavelength spatial control of both polarization and phase. Several diffractive optical elements, including high numerical aperture lenses and simultaneous phase and polarization controllers have recently been demonstrated with high efficiencies~\cite{Arbabi2015NatCommun,Arbabi2015NatNano}. These devices are based on subwavelength arrays of high refractive index dielectric nano-resonators (scatterers) with different geometries, fabricated on a planar substrate. Scatterers with various geometries impart different phases to the transmitted light, shaping its wavefront to the desired form.

One main drawback of almost all of metasurface devices, particularly the ones with spatially varying phase profiles like lenses and gratings, is their chromatic aberration: their performance changes as the wavelength is varied~\cite{Young1972JOSA,Born1999,Saleh2013}. Refractive optical elements also suffer from chromatic aberrations; however, their chromatic aberrations, which stem from material dispersion, are substantially smaller than those of the diffractive elements~\cite{Born1999,Saleh2013}. An ideal refractive lens made of a dispersionless material will show no chromatic aberration. On the other hand, the chromatic aberration of diffractive elements mainly comes from the geometrical arrangement of the device. Early efforts focused on making achromatic diffractive lenses by cascading them in the form of doublets and triplets~\cite{Latta1972AppOpt,Bennett1976AppOpt,Sweatt1977AppOpt,Weingartner1982OpticaActa}, but it was later shown that it is fundamentally impossible to make a converging achromatic lens which has a paraxial solution (i.e. is suitable for imaging) by only using diffractive elements~\cite{Buralli1989JOSAA}. Although diffractive-refractive combinations have successfully been implemented to reduce chromatic aberrations, they are mostly useful in deep UV and X-ray wavelengths where materials are significantly more dispersive~\cite{Swanson1989,Wang2003Nature}. More recently, wavelength and polarization selectivity of metallic meta-atoms have been used to fabricate a Fresnel zone plate lens that operates at two distinct wavelengths with different orthogonal polarizations~\cite{Eisenbach2015OptExp}. Besides undesired multi-focus property of Fresnel zone plates and efficiency limitations of metallic metasurfaces~\cite{Swanson1989,Aieta2012NanoLett,Arbabi2014arXiv}, the structure works only with different polarizations at the two wavelengths. The large phase dispersion of dielectric meta-atoms with multiple resonances has also been exploited to compensate for the phase dispersion of metasurfaces at three wavelengths~\cite{Aieta2015Science,Khorasaninejad2015NanoLett}, but the cylindrical lens demonstrated with this technique is polarization dependent and has low numerical aperture and efficiency. Multiwavelength metasurfaces based on elliptical apertures in metallic films are demonstrated in~\cite{Zhao2015SciRep}, but they are also polarization dependent and have a multi-focus performance. An achromatic metasurface design is proposed in~\cite{Cheng2015JOSAB} based on the idea of dispersionless meta-atoms (i.e. meta-atoms that impart constant delays). Unfortunately, this idea only works for metasurface lenses with one Fresnel zone limiting the size and numerical aperture of the lenses. For a typical lens with tens of Fresnel zones, dispersionless meta-atoms will not reduce the chromatic dispersion as we will shortly discuss. In the following we briefly discuss the reason for chromatic dispersion of metasurface lenses, and then propose a method for correcting this dispersion at distinct wavelengths. We also present experimental results demonstrating corrected behavior of a lens realized using the proposed method.

\begin{figure*}[htp]

\includegraphics{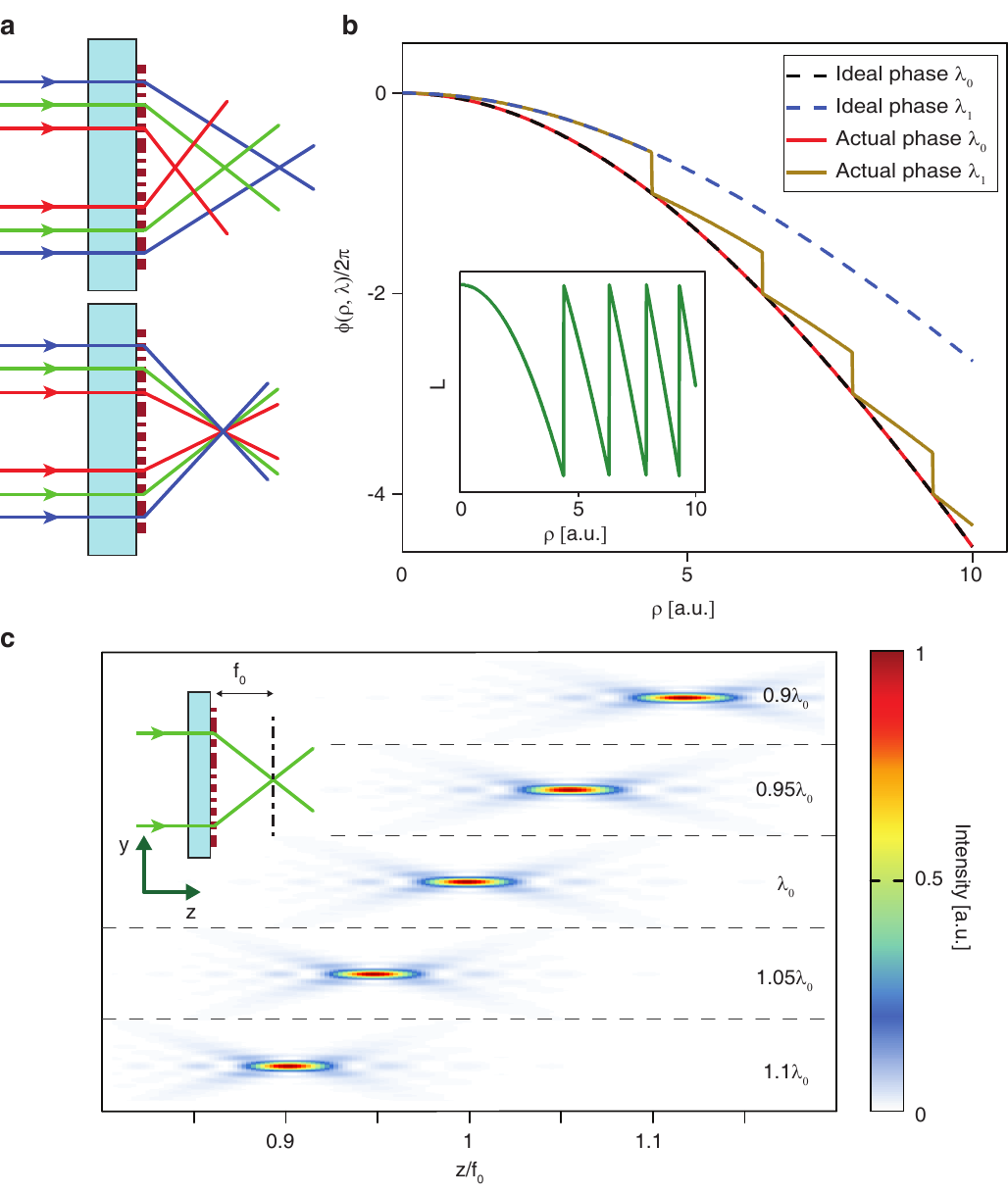}
\caption{\textbf{Chromatic dispersion of metasurface lenses.} \textbf{a}, Schematic illustration of a typical metasurface lens focusing light of different wavelengths to different focal distances (top), and a metasurface lens corrected to focus light at specific different wavelengths to the same focal distance (bottom). \textbf{b}, The phase profile of a hypothetical aspherical metasurface lens at the design wavelength $\lambda_0$ and a different wavelength $\lambda_1$ as a function of the distance to the center of the lens ($\rho$). (Inset) Plot of the parameter of the meta-atoms controlling phase (named $L$). \textbf{c}, Intensity of light at different wavelengths in the axial plane after passing through the lens showing considerable chromatic dispersion rising from phase jumps at the boundaries between different Fresnel zones.}
\end{figure*}

In diffractive lenses, chromatic dispersion mainly manifests itself through a significant change of focal length as a function of wavelength~\cite{Swanson1989}. This change is schematically shown in Fig. 1a, along with a metasurface lens corrected to have the same focal distance at a few wavelengths. To better understand the underlying reasons for this chromatic dispersion, we consider a hypothetical aspherical metasurface lens. The lens is composed of different meta-atoms which locally modify the phase of the transmitted light to generate the desired wavefront. We assume that the meta-atoms are dispersionless in the sense that their associated phase changes with wavelength as $\phi(\lambda)={2\pi L}/{\lambda}$ like a piece of dielectric with a constant refractive index. Here $\lambda$ is the wavelength and $L$ is an effective parameter associated with the meta-atoms that controls the phase ($L$ can be an actual physical parameter or a function of physical parameters of the meta-atoms). We assume that the full 2$\pi$ phase needed for the lens is covered using different meta-atoms with different values of $L$. The lens is designed to focus light at $\lambda_0$ (Fig. 1b) to a focal distance $f_0$, and it's phase profile in all Fresnel zones matches the ideal phase profile at this wavelength. Because of the specific wavelength dependence of the dispersionless meta-atoms (i.e. proportionality to 1/$\lambda$ ), at a different wavelength ($\lambda_1$) the phase profile of the lens in the first Fresnel zone follows the desired ideal profile needed to maintain the same focal distance (Fig. 1b). However, outside the first Fresnel zone, the actual phase profile of the lens deviates substantially from the desired phase profile. Due to the jumps at the boundaries between the Fresnel zones, the actual phase of the lens at $\lambda_1$ is closer to the ideal phase profile at $\lambda_0$ than the desired phase profile at $\lambda_1$. In the inset of Fig. 1b the effective meta-atom parameter $L$ is plotted as a function of distance to the center of the lens $\rho$. The jumps in $L$ coincide with the jumps in the phase profile at $\lambda_1$. In Fig. 1c, the simulated intensity profile of the same hypothetical lens is plotted at a few wavelengths close to $\lambda_0$. The focal distance changes approximately proportional to $1/\lambda$. This wavelength dependence is also observed in Fresnel zone plates~\cite{Swanson1989}, and  for lenses with wavelength independent phase profiles~\cite{Born1999, Saleh2013} (the  $1/\lambda$ dependence is exact in the paraxial limit, and approximate in general). This behavior confirms the previous observation that the phase profile of the lens at other wavelengths approximately follows the phase profile at the design wavelength. Therefore, the chromatic dispersion of metasurface lenses mainly stems from wrapping the phase, and the dependence of the phase on only one effective parameter (e.g. $L$) whose value undergoes sudden changes at the zone boundaries. As we show in the following, using two parameters to control metasurface phase at two wavelengths can resolve this issue, and enable lenses with the same focal lengths at two different wavelengths.

\begin{figure*}[htp]
\centering
\includegraphics[width=2\columnwidth]{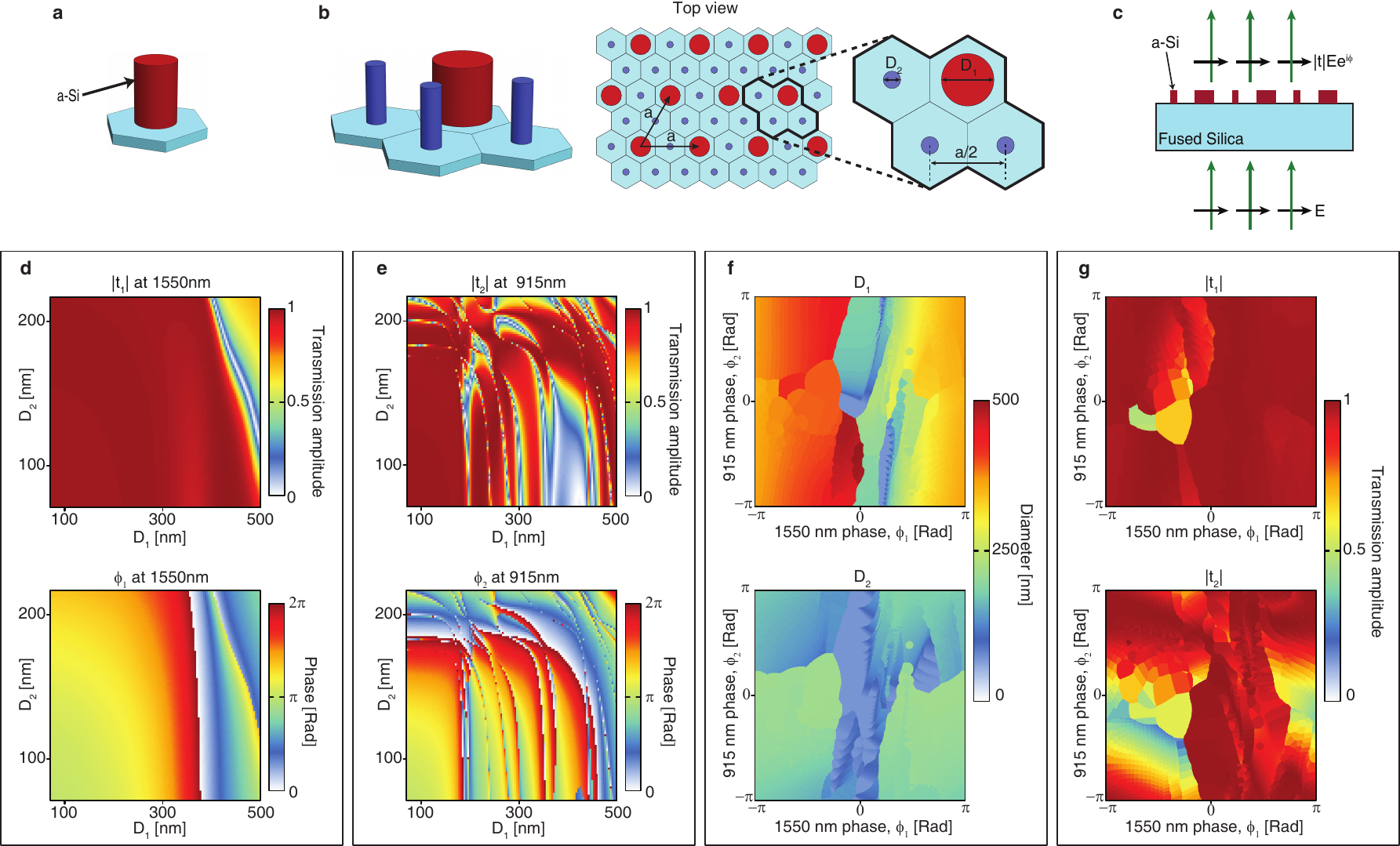}
\caption{\textbf{Meta-molecule design and its transmission characteristics.} \textbf{a}, The single scattering element composed of an a-Si nano-post on a fused silica substrate. \textbf{b}, The unit cell composed of four scattering elements that provide more control parameters for the scattering phase (left). The meta-molecules are placed on a hexagonal lattice with lattice constant $a$ (middle). Top view of a single meta-molecule (right). \textbf{c}, Schematic of the structure simulated for finding the transmission coefficient of the metasurface. \textbf{d} and \textbf{e}, Transmission amplitude (top) and phase (bottom) as a function of the two diameters in the unit cell for 1550 nm and 915 nm. \textbf{f}, Selected values of $D_1$ (top) and $D_2$ (bottom) as functions of phases at 1550 nm ($\phi_1$) and 950 nm ($\phi_2$).}
\end{figure*}

The metasurface platform we use in this work is based on amorphous silicon (a-Si) nano-posts on a fused silica substrate (Fig. 2a). The nano-posts are placed on the vertices of a hexagonal lattice, and locally sample the phase to generate the desired phase profile~\cite{Arbabi2015NatCommun}. For a fixed height, the transmission phase of a nano-post can be controlled by varying its diameter. The posts’ height can be chosen such that at a certain wavelength the whole 2$\pi$ phase shift is covered, while keeping the transmission amplitude high. To design a metasurface that works at two different wavelengths, a unit cell consisting of four different nano-posts (Fig. 2b) was chosen because it has more parameters to control the phases at two wavelengths almost independently. As molecules consisting of multiple atoms form the units for more complex materials, we call these unit cells with multiple meta-atoms \textit{meta-molecules}. The meta-molecules can also form a periodic lattice (in this case hexagonal), and effectively sample the desired phase profiles simultaneously at two wavelengths. The lattice is subwavelength at both wavelengths of interest; therefore, the non-zero diffraction orders are not excited. In general, the four nano-posts can each have different diameters and distances from each other. However, to make the design process more tractable, we choose three of the four nano-posts with the same diameter $D_1$ and the fourth post with diameter $D_2$, and place them in the centers of the hexagons at a distance $a/2$ (as shown in Fig. 2b).  Therefore, each meta-molecule has two parameters, $D_1$ and $D_2$, to control the phases at two wavelengths. For this demonstration, we choose two wavelengths of 1550 nm and 915 nm, because of the availability of lasers at these wavelengths. A periodic array of meta-molecules was simulated to find the transmission amplitude and phase as shown in Fig. 2c (see methods for simulation details). The simulated transmission amplitude and phase for 1550 nm ($|t_1|$ and $\phi_1$) and 915 nm ($|t_2|$ and $\phi_2$) are plotted as functions of $D_1$ and $D_2$ in Fig. 2d and 2e. In these simulations the lattice constant ($a$) was set to 720 nm and the posts were 718 nm tall. Since the two wavelengths are not close, the ranges of $D_1$ and $D_2$ must be very different in order to properly control the phases at 1550 nm and 915 nm. For each desired combination of the phases $\phi_1$ and $\phi_2$ in the $(-\pi, \pi)$ range at the two wavelengths, there is a corresponding $D_1$ and $D_2$ pair that minimizes the total transmission error which is defined as $\epsilon=|\mathrm{exp}(i\phi_1)-t_1|^2+|\mathrm{exp}(i\phi_2)-t_2|^2$. These pairs are plotted in Fig. 2f as a function of $\phi_1$ and $\phi_2$. Using the complex transmission coefficients (i.e. $t_1$ and $t_2$) in error calculations results in automatically avoiding resonance areas where the phase might be close to the desired value, but transmission is low. The corresponding transmission amplitudes for the chosen meta-molecules are plotted in Fig. 2g, and show this automatic avoidance of low transmission meta-molecules. In the lens design process, the desired transmission phases of the lens are sampled at the lattice points at both wavelengths resulting in a $(\phi_1, \phi_2)$ pair at each lattice site. Using the plots in Fig. 2f, values of the two post diameters are found for each lattice point. Geometrically, the values of the two diameters are limited by $D_1 + D_2 < a$. Besides, we set a minimum value of 50 nm for the gaps between the posts to facilitate the metasurface fabrication.

\begin{figure*}[htp]
\centering
\includegraphics{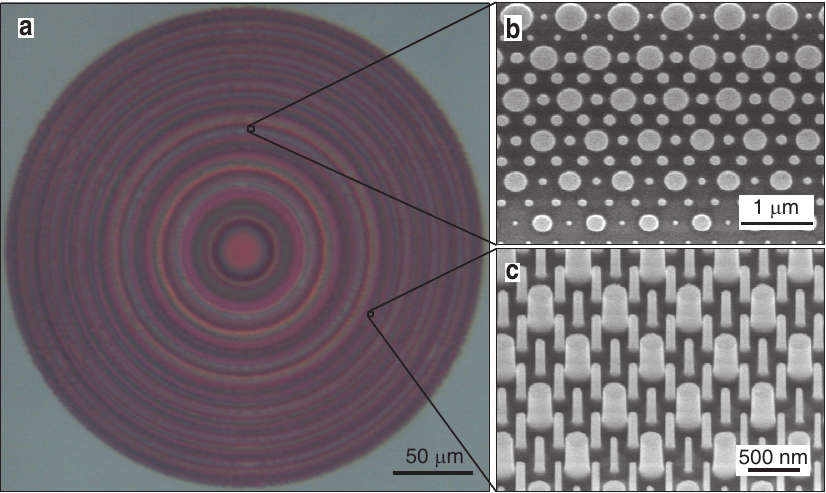}
\caption{\textbf{Fabricated device images.} \textbf{a}, Optical microscope image of the fabricated device. \textbf{b} and \textbf{c}, Scanning electron micrographs of parts of the fabricated device from top \textbf{b} and with a 30 degrees tilt \textbf{c}. }
\end{figure*}

A double wavelength aspherical lens was designed using the proposed platform to operate at both 1550 nm and 915 nm. The lens has a diameter of 300 $\mu$m and focuses the light emitted from single mode fibers at each wavelength to a focal plane 400 $\mu$m away from the lens surface (the corresponding paraxial focal distance is 286 $\mu$m, thus the numerical aperture is 0.46). The lens was fabricated using standard nanofabrication techniques: a 718-nm-thick layer of a-Si was deposited on a fused silica substrate, the lens pattern was generated using electron beam lithography and transferred to the a-Si layer using aluminum oxide as a hard mask (see methods for more details on fabrication). Optical and scanning electron microscope images of the lens and nano-posts are shown in Fig. 3. For characterization, the fabricated metasurface lens was illuminated by light emitted from the end facet of a single mode fiber, and the transmitted light intensity was imaged at different distances from the lens using a custom built microscope (see methods and Supplementary Information Fig. S1 for measurement details). Measurement results for both wavelengths are plotted in Fig. 4a-4d. Figures 4a and 4b show the intensity profiles in the focal plane measured at 915 nm and 1550 nm, respectively. The measured full width at half maximum (FWHM) is 1.9 $\mu$m at 915 nm, and 2.9 $\mu$m at 1550 nm. The intensity measured at the two axial plane cross sections are plotted in Fig. 4c and 4d for the two wavelengths. A nearly diffraction limited focus is observed in the measurements, and no other secondary focal points with comparable intensity is seen. To confirm the diffraction limited behavior, a perfect phase mask was simulated using the same illumination as the measurements. The simulated FWHM's were 1.6 $\mu$m and 2.75 $\mu$m for 915 nm and 1550 nm respectively (see methods for details on the simulation). Focusing efficiencies of 22$\%$ and 65$\%$ were measured for 915 nm and 1550 nm, respectively. Focusing efficiency is defined as the ratio of the power passing through a 20-$\mu$m-diameter disk around the focus to the total power incident on the lens. Anther lens with a longer focal distance of 1000 $\mu$m (thus a lower NA of 0.29) was fabricated and measured with the same platform and method. Measurement results for those devices are presented in Supplementary Information Fig. S2. Slightly higher focusing efficiencies of 25$\%$ and 72$\%$ were measured at 915 nm and 1550 nm for those devices. For comparison, a lens designed with the same method and based on the same metasurface platform is simulated using finite difference time domain (FDTD) method with a freely available software (MEEP)~\cite{Oskooi2010CompPhys}. To reduce the computational cost, the simulated lens is four times smaller and focuses the light at 100 $\mu$m distance. Because of the equal numerical apertures of the simulated and fabricated devices, the focal intensity distributions and the focal depths are comparable. The simulation results are shown in Fig. 4e-4h. Figures 4e and 4f show the simulated focal plane intensity of the lens at 915 nm and 1550 nm, respectively. The simulated FWHM is 1.9 $\mu$m at 915 nm and 3 $\mu$m at 1550 nm, both of which are in accordance with their corresponding measured values. Also, the simulated intensity distributions in the axial cross section planes, which are shown in Fig. 4g and 4h, demonstrate only one strong focal point. The focusing efficiency was found to be 32$\%$ at 915 nm, and 73$\%$ at 1550 nm. We attribute the difference in the simulated and measured efficiencies to fabrication imperfections and measurement artifacts (see methods for details about measurements).

\begin{figure*}[htp]
\centering
\includegraphics{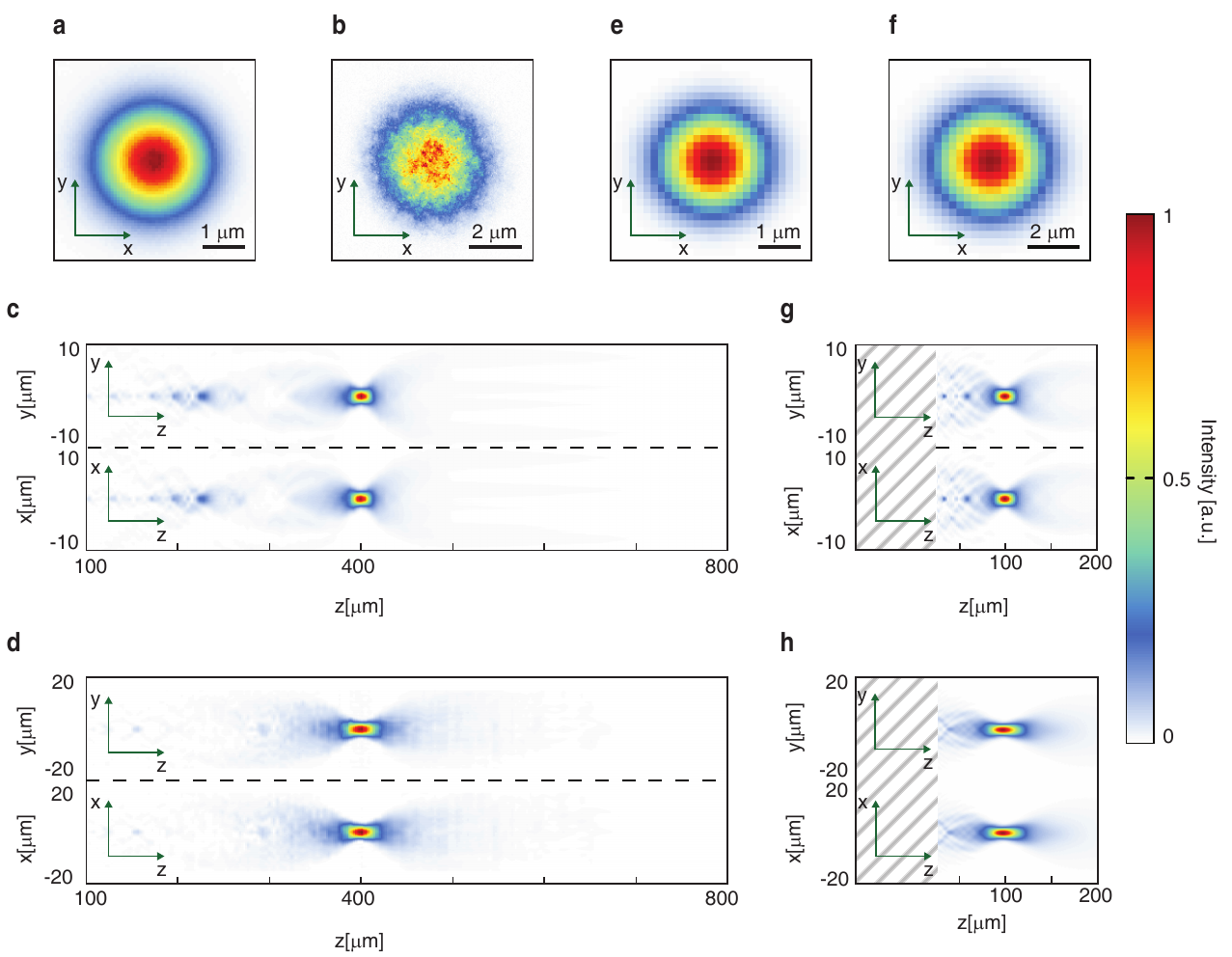}
\caption{\textbf{Measurement and simulation results of the double-wavelength lenses.} \textbf{a and b}, Measured intensity in the focal plane of the double wavelength lens (400 $\mu$m focal length, 300 $\mu$m diameter) at 915 nm \textbf{a} and 1550 nm \textbf{b}. \textbf{c}, Intensity measured in the axial planes of the lens for 915 nm. \textbf{d} The same axial plots for 1550 nm. \textbf{e} Simulated focal plane intensity of a lens with the same numerical aperture as the one shown in \textbf{a}-\textbf{d} but with a dimeter of 75 $\mu$m at wavelength of 915 nm, and \textbf{f} at 1550 nm. \textbf{g} and \textbf{h}, Simulated intensity profiles in the axial planes at 915 nm and 1550 nm, respectively, calculated for the same lens described in \textbf{e}.}
\end{figure*}

The efficiency at 915 nm is found to be lower than what expected both in measurement and FDTD simulation. While the average power transmission of the selected meta-molecules is about 73$\%$ as calculated from Fig. 2g, the simulated focusing efficiency is about 32$\%$. To better understand the reasons for this difference, two blazed gratings with different angles were designed and simulated for both wavelengths using the same meta-molecules (see Supplementary Information Section 1 and Fig. S3). It is observed that for the gratings (that are aperiodic), a significant portion of the power is diffracted to other angles both in reflection and transmission. Besides, the power lost into diffractions to other angles is higher for the grating with larger deflection angle. The main reason for the large power loss to other angles is the the relatively large lattice constant. The chosen lattice constant of  $a=$ 720 nm is just slightly smaller than 727 nm, the lattice constant at which the first order diffracted light starts to propagate in the glass substrate for a perfectly periodic structure. Thus, even a small deviation from perfect periodicity can result in light diffracted to propagating orders. Besides, the lower transmission of some meta-molecules reduces the purity of the plane wave wavefronts diffracted to the design angle. Furthermore, the desired phase profile of high numerical aperture lenses cannot be sampled at high enough resolution using large lattice constants. Therefore, as shown in this work, a lens with a lower numerical aperture has a higher efficiency. There are a few methods to increase the efficiency of the lenses at 915 nm: The lattice constant is bound by the geometrical and fabrication constraint: $D_1 + D_2 + 50\mathrm{nm} < a$, hence the smallest value of $D_1 + D_2$ that gives full phase coverage at the longer wavelength sets the lower bound for the lattice constant. This limit can usually be decreased by using taller posts, however, that would result in a high sensitivity to fabrication errors at the shorter wavelength. Thus, a compromise should be made here, and higher efficiency designs might be possible by more optimum selections of the posts’ height and the lattice constant.  The lattice constant can also be smaller if less than the full 2$\pi$ phase shift is used at 1550 nm (thus lower efficiency at 1550 nm). In addition, as explained earlier, in minimizing the total transmission error equal weights are used for 915 nm and 1550 nm. A higher weight for 915 nm might result in higher efficiency at this wavelength, probably at the expense of 1550 nm efficiency. For instance, if we optimize the lens only for operation at 915 nm, devices with efficiencies as high as 80$\%$ are possible~\cite{Arbabi2015NatCommun}.

The approach presented here cannot be directly used to correct for chromatic dispersion over a continuous bandwidth; the multiwavelength lenses still have chromatic dispersion much like normal metasurface lenses in narrow bandwidths around the corrected wavelengths. For achieving zero chromatic dispersion over a narrow bandwidth, the meta-atoms should independently control the phase at two very close wavelengths. High quality factor resonances must be present for the meta-atom phase to change rapidly over a narrow bandwidth, and such resonances will result in high sensitivities to fabrication errors that would make the metasurface impractical.

The meta-molecule platform, used here to correct for chromatic aberration at specific wavelengths, can also be used for applications where different functionalities at different wavelengths are desired. For instance, it can be used to implement a lens with two given focal distances at two wavelengths, or a lens converging at one wavelength and diverging at the other. Multi-wavelength operation is necessary in various microscopy applications where fluorescence is excited at one wavelength and collected at another. In this work we have used only two of the degrees of freedom of the meta-molecules, but increased functionalities at more than two wavelengths can be achieved by making use of all the degrees of freedom. Operation at more than two wavelengths enables applications in color display technologies or more complex fluorescence imaging techniques.

\section*{Methods}

\noindent\textbf{Simulation.} To find the transmission amplitude and phase of a multi-element metasurface, the rigorous coupled wave analysis technique was used~\cite{Liu2012CompPhys}. A normally incident plane wave at each wavelength was used as the excitation, and the amplitude and phase of the transmitted wave were extracted. Since the lattice is subwavelength for normal incidence at both wavelengths, only the zeroth order diffracted light is nonzero. This justifies the use of only one transmission value at each wavelength to describe the behavior of meta-atoms. The lattice constant was chosen as 720 nm, and the a-Si posts were 718-nm tall. Refractive indices of 3.56 and 3.43 were assumed for a-Si at 915 nm and 1550 nm, respectively.

The paraxial focal distance of the two lenses were calculated to be 286 $\mu$m and 495 $\mu$m for the lenses that focus light from the fiber to 400 $\mu$m and 1000 $\mu$m respectively, by fitting a parabola to the phase profiles of the lenses. For a fitted parabola $y=\alpha x^2$, the paraxial focal distance can be calculated using $f=2\pi / 2 \alpha \lambda$. The corresponding numerical apertures can then found to be 0.46 and 0.29 for the two lenses.

The perfect phase mask (that also served as the goal phase profile for the designed devices) was calculated from the illuminating field and the aspherical desired phase profile using the method described in supplementary information of~\cite{Arbabi2015NatCommun}. The illuminating field was calculated by propagating the output fields of single mode fibers at each wavelength using plane wave expansion (PWE) method up to the metasurface layer. The perfect phase mask was then applied to the field, and the result was propagated using the PWE method to the focal point. The diffraction limited FWHM was then calculated from the intensity profile at the focal plane.

Full wave simulation of a full lens was done using finite difference time domain method (FDTD) in MEEP~\cite{Oskooi2010CompPhys}. A lens with a diameter of 75 $\mu$m and a focal length of 100 $\mu$m was designed with the same method as the fabricated device. to the lens focuses the light emitted from a single mode fiber (with mode diameters of 10.4 $\mu$m at 1550 nm and 6 $\mu$m at 915 nm) placed 150 $\mu$m away from a 125 $\mu$m thick fused silica substrate (all of the geometrical dimensions were chosen 4 times smaller than the values for the experimentally measured device). The distances to fibers were chosen such that more than 99$\%$ of the total power emitted by the fiber passes through the lens aperture. At both wavelengths, the light from the fibers was propagated through air, air-glass interface, and through glass up to a plane about a wavelength before the metasurface using a plane wave expansion (PWE) code. Electric and magnetic field distributions at this plane were used as sources for FDTD simulation of the lenses, and fields were calculated at about a wavelength after the metasurface using MEEP. The PWE code was used again to further propagate these fields to the focal plane and beyond (Fig. 3e-3h). The focusing efficiencies were calculated by dividing the power in a 20-$\mu$m-diameter disk around the focus, to the total power incident on the lens.

\noindent\textbf{Sample fabrication.} A 718-nm-thick hydrogenated a-Si layer was deposited on a fused silica substrate using the plasma enhanced chemical vapor deposition (PECVD) technique with a 5\% mixture of silane in argon at $200\,^{\circ}{\rm C}$. A Vistec EBPG5000+ electron beam lithography system was used to define the metasurface pattern in the ZEP-520A positive resist ($\sim$300 nm, spin coated at 5000 rpm for 1 min). The pattern was developed in a resist developer for 3 minutes (ZED-N50 from Zeon Chemicals). An approximately 100-nm-thick aluminum oxide layer was deposited on the sample using electron beam evaporation, and was lifted off reversing the pattern. The patterned aluminum oxide hard mask was then used to dry etch the a-Si layer in a 3:1 mixture of $\mathrm{SF_6}$ and $\mathrm{C_4F_8}$ plasma. After etching, the mask was removed using a 1:1 solution of ammonium hydroxide and hydrogen peroxide at $80^{\circ}$ C.

\noindent\textbf{Measurement procedure.} Devices were measured using a fiber placed $\sim$1100 $\mu$m away from the metasurface (500 $\mu$m substrate thickness plus 600 $\mu$m distance between the fiber and the substrate), and a custom built microscope with $\sim$100X magnification (Supplementary Information Fig. S2). At 915 nm, a fiber coupled semiconductor laser with a single mode fiber with an angled polished connector was used for illumination. Fiber tip angle was adjusted to correct for the angled connector cut. A 100X objective lens (Olympus UMPlanFl, NA=0.95) and a tube lens (Thorlabs AC254-200-B-ML) with a focal distance of 20 cm were used to image intensity at planes of interest to a CCD camera (CoolSNAP K4, Photometrics). A calibration sample with known feature sizes was measured to find the pixel-size transferred to the object plane. The objective was moved with a translation stage to image different planes around the focus. The plotted axial plane intensities are upsampled 2:1 in the axial direction (4 $\mu$m adjacent measurement planes distance to 2 $\mu$m) to achieve a smoother graph. For focusing efficiency measurement at 915 nm, a 20-$\mu$m-diameter pinhole was placed in the focal plane of the metasurface lens to only let the focused light pass through. The pinhole was made by wet etching a 20 $\mu$m hole in a thick layer of chrome deposited on a fused silica substrate. A power meter (Thorlabs PM100D with photodetector head Thorlabs S122C) was then used to measure the power after the pinhole, and the output power of the fiber. The efficiency was calculated as the ratio of these two powers.  The reported measured efficiency is therefore a lower bound on the actual efficiency as it does not include reflection from the substrate, and two reflections from the two sides of the pinhole glass substrate.
A similar setup was used for measurements at 1550: a tunable 1550 nm laser (Photonetics Tunics-Plus) was used with a single mode fiber for illumination. The same 100X objective was used with a 20 cm tube lens (Thorlabs AC254-200-C-ML) to image the intensity in the object plane to a camera (Digital CamIR 1550 by Applied Scintillation Technologies). The camera has a significantly non-uniform sensitivity for different pixels which leads to high noise level of the images captured by the camera (as seen in Fig. 3b). The nonphysical high frequency noise of the images (noise with frequencies higher than twice the free space propagation constant) was removed numerically to reduce the noise in the axial intensity patterns. The intensity pattern was also upsampled in the axial direction from the actual 4 $\mu$m distance between adjacent measurement planes, to 2 $\mu$m to achieve a smoother intensity profile.  To find the focused power, the focal plane of the lens was imaged using the microscope to a photodetector. A 2 mm iris in the image plane (corresponding to 20 $\mu m$ in the object plane) was used to limit the light reaching the photodetector. The input power was measured by imaging the fiber facet to the photodetector using the same setup and without the iris. The efficiency was obtained by dividing the focused power by the input power.

\noindent\textbf{Supplementary Information.} Discussion of 915 nm efficiencies, measurement setup figure, measurement results for the lower NA lens, simulation results of two 915 nm blazed gratings.

\section*{Acknowledgement}
This work was supported by Samsung Electronics. A.A and Y.H were also supported by DARPA, and S.K was supported as part of the Department of Energy (DOE) ``Light-Material Interactions in Energy Conversion” Energy Frontier Research Center under grant no. DE-SC0001293. The device nanofabrication was performed at the Kavli Nanoscience Institute at Caltech.

\textbf{Author contributions}\\
E. A., A. A., and A. F. conceived the experiments. E. A., A. A., S. K, and Y. H. performed the simulations and fabricated the devices. E. A. and A. A. performed the measurements, and analyzed the data. E. A., A. A., S. K., and A. F. co-wrote the manuscript. All authors discussed the results and commented on the manuscript.

\clearpage

\bibliographystyle{naturemag_noURL}

\begin{thebibliography}{10}
\expandafter\ifx\csname url\endcsname\relax
  \def\url#1{\texttt{#1}}\fi
\expandafter\ifx\csname urlprefix\endcsname\relax\def\urlprefix{URL }\fi
\providecommand{\bibinfo}[2]{#2}
\providecommand{\eprint}[2][]{\url{#2}}

\bibitem{Astilean1998OptLett}
\bibinfo{author}{Astilean, S.}, \bibinfo{author}{Lalanne, P.},
  \bibinfo{author}{Chavel, P.}, \bibinfo{author}{Cambril, E.} \&
  \bibinfo{author}{Launois, H.}
\newblock \bibinfo{title}{High-efficiency subwavelength diffractive element
  patterned in a high-refractive-index material for 633 nm}.
\newblock \emph{\bibinfo{journal}{Opt. Lett.}} \textbf{\bibinfo{volume}{23}},
  \bibinfo{pages}{552--554} (\bibinfo{year}{1998}).

\bibitem{Lalanne1999JOSAA}
\bibinfo{author}{Lalanne, P.}, \bibinfo{author}{Astilean, S.},
  \bibinfo{author}{Chavel, P.}, \bibinfo{author}{Cambril, E.} \&
  \bibinfo{author}{Launois, H.}
\newblock \bibinfo{title}{Design and fabrication of blazed binary diffractive
  elements with sampling periods smaller than the structural cutoff}.
\newblock \emph{\bibinfo{journal}{J. Opt. Soc. Am. A}}
  \textbf{\bibinfo{volume}{16}}, \bibinfo{pages}{1143--1156}
  (\bibinfo{year}{1999}).

\bibitem{Fattal2010NatPhoton}
\bibinfo{author}{Fattal, D.}, \bibinfo{author}{Li, J.}, \bibinfo{author}{Peng,
  Z.}, \bibinfo{author}{Fiorentino, M.} \& \bibinfo{author}{Beausoleil, R.~G.}
\newblock \bibinfo{title}{Flat dielectric grating reflectors with focusing
  abilities}.
\newblock \emph{\bibinfo{journal}{Nat. Photonics}}
  \textbf{\bibinfo{volume}{4}}, \bibinfo{pages}{466--470}
  (\bibinfo{year}{2010}).

\bibitem{Lin2014Science}
\bibinfo{author}{Lin, D.}, \bibinfo{author}{Fan, P.}, \bibinfo{author}{Hasman,
  E.} \& \bibinfo{author}{Brongersma, M.~L.}
\newblock \bibinfo{title}{Dielectric gradient metasurface optical elements}.
\newblock \emph{\bibinfo{journal}{Science}} \textbf{\bibinfo{volume}{345}},
  \bibinfo{pages}{298--302} (\bibinfo{year}{2014}).

\bibitem{Kildishev2013Science}
\bibinfo{author}{Kildishev, A.~V.}, \bibinfo{author}{Boltasseva, A.} \&
  \bibinfo{author}{Shalaev, V.~M.}
\newblock \bibinfo{title}{Planar photonics with metasurfaces}.
\newblock \emph{\bibinfo{journal}{Science}} \textbf{\bibinfo{volume}{339}}
  (\bibinfo{year}{2013}).

\bibitem{Yu2014NatMater}
\bibinfo{author}{Yu, N.} \& \bibinfo{author}{Capasso, F.}
\newblock \bibinfo{title}{Flat optics with designer metasurfaces}.
\newblock \emph{\bibinfo{journal}{Nat. Mater.}} \textbf{\bibinfo{volume}{13}},
  \bibinfo{pages}{139--150} (\bibinfo{year}{2014}).

\bibitem{Vo2014IEEEPhotonTechLett}
\bibinfo{author}{Vo, S.} \emph{et~al.}
\newblock \bibinfo{title}{Sub-wavelength grating lenses with a twist}.
\newblock \emph{\bibinfo{journal}{IEEE Photonics Technol. Lett.}}
  \textbf{\bibinfo{volume}{26}}, \bibinfo{pages}{1375--1378}
  (\bibinfo{year}{2014}).

\bibitem{Arbabi2015NatCommun}
\bibinfo{author}{Arbabi, A.}, \bibinfo{author}{Horie, Y.},
  \bibinfo{author}{Ball, A.~J.}, \bibinfo{author}{Bagheri, M.} \&
  \bibinfo{author}{Faraon, A.}
\newblock \bibinfo{title}{Subwavelength-thick lenses with high numerical
  apertures and large efficiency based on high-contrast transmitarrays}.
\newblock \emph{\bibinfo{journal}{Nat. Commun.}} \textbf{\bibinfo{volume}{6}}
  (\bibinfo{year}{2015}).

\bibitem{Arbabi2015NatNano}
\bibinfo{author}{Arbabi, A.}, \bibinfo{author}{Horie, Y.},
  \bibinfo{author}{Bagheri, M.} \& \bibinfo{author}{Faraon, A.}
\newblock \bibinfo{title}{Dielectric metasurfaces for complete control of phase
  and polarization with subwavelength spatial resolution and high
  transmission}.
\newblock \emph{\bibinfo{journal}{Nat. Nanotechnol.}}
  \textbf{\bibinfo{volume}{10}}, \bibinfo{pages}{937--943}
  (\bibinfo{year}{2015}).

\bibitem{Young1972JOSA}
\bibinfo{author}{Young, M.}
\newblock \bibinfo{title}{Zone plates and their aberrations}.
\newblock \emph{\bibinfo{journal}{J. Opt. Soc. Am.}}
  \textbf{\bibinfo{volume}{62}}, \bibinfo{pages}{972--976}
  (\bibinfo{year}{1972}).

\bibitem{Born1999}
\bibinfo{author}{Born, M.}, \bibinfo{author}{Wolf, E.} \&
  \bibinfo{author}{Bhatia, A.}
\newblock \emph{\bibinfo{title}{Principles of Optics: Electromagnetic Theory of
  Propagation, Interference and Diffraction of Light}}
  (\bibinfo{publisher}{Cambridge University Press}, \bibinfo{year}{1999}).

\bibitem{Saleh2013}
\bibinfo{author}{Saleh, B.} \& \bibinfo{author}{Teich, M.}
\newblock \emph{\bibinfo{title}{Fundamentals of Photonics}}
  (\bibinfo{publisher}{Wiley}, \bibinfo{year}{2013}).

\bibitem{Latta1972AppOpt}
\bibinfo{author}{Latta, J.~N.}
\newblock \bibinfo{title}{Analysis of multiple hologram optical elements with
  low dispersion and low aberrations}.
\newblock \emph{\bibinfo{journal}{Appl. Opt.}} \textbf{\bibinfo{volume}{11}},
  \bibinfo{pages}{1686--1696} (\bibinfo{year}{1972}).

\bibitem{Bennett1976AppOpt}
\bibinfo{author}{Bennett, S.~J.}
\newblock \bibinfo{title}{Achromatic combinations of hologram optical
  elements}.
\newblock \emph{\bibinfo{journal}{Appl. Opt.}} \textbf{\bibinfo{volume}{15}},
  \bibinfo{pages}{542--545} (\bibinfo{year}{1976}).

\bibitem{Sweatt1977AppOpt}
\bibinfo{author}{Sweatt, W.}
\newblock \bibinfo{title}{Achromatic triplet using holographic optical
  elements}.
\newblock \emph{\bibinfo{journal}{Appl. Opt.}} \textbf{\bibinfo{volume}{16}},
  \bibinfo{pages}{1390--1391} (\bibinfo{year}{1977}).

\bibitem{Weingartner1982OpticaActa}
\bibinfo{author}{Weingärtner, I.} \& \bibinfo{author}{Rosenbruch, K.~J.}
\newblock \bibinfo{title}{Chromatic correction of two- and three-element
  holographic imaging systems}.
\newblock \emph{\bibinfo{journal}{Opt. Acta}} \textbf{\bibinfo{volume}{29}},
  \bibinfo{pages}{519--529} (\bibinfo{year}{1982}).

\bibitem{Buralli1989JOSAA}
\bibinfo{author}{Buralli, D.~A.} \& \bibinfo{author}{Rogers, J.~R.}
\newblock \bibinfo{title}{Some fundamental limitations of achromatic
  holographic systems}.
\newblock \emph{\bibinfo{journal}{J. Opt. Soc. Am. A}}
  \textbf{\bibinfo{volume}{6}}, \bibinfo{pages}{1863--1868}
  (\bibinfo{year}{1989}).

\bibitem{Swanson1989}
\bibinfo{author}{Swanson, G.~J.}
\newblock \bibinfo{title}{Binary optics technology: the theory and design of
  multi-level diffractive optical elements}.
\newblock \bibinfo{type}{Tech. Rep.}, \bibinfo{institution}{DTIC Document}
  (\bibinfo{year}{1989}).

\bibitem{Wang2003Nature}
\bibinfo{author}{Wang, Y.}, \bibinfo{author}{Yun, W.} \&
  \bibinfo{author}{Jacobsen, C.}
\newblock \bibinfo{title}{Achromatic fresnel optics for wideband
  extreme-ultraviolet and x-ray imaging}.
\newblock \emph{\bibinfo{journal}{Nature}} \textbf{\bibinfo{volume}{424}},
  \bibinfo{pages}{50--53} (\bibinfo{year}{2003}).

\bibitem{Eisenbach2015OptExp}
\bibinfo{author}{Eisenbach, O.}, \bibinfo{author}{Avayu, O.},
  \bibinfo{author}{Ditcovski, R.} \& \bibinfo{author}{Ellenbogen, T.}
\newblock \bibinfo{title}{Metasurfaces based dual wavelength diffractive
  lenses}.
\newblock \emph{\bibinfo{journal}{Opt. Express}} \textbf{\bibinfo{volume}{23}},
  \bibinfo{pages}{3928--3936} (\bibinfo{year}{2015}).

\bibitem{Aieta2012NanoLett}
\bibinfo{author}{Aieta, F.} \emph{et~al.}
\newblock \bibinfo{title}{Aberration-free ultrathin flat lenses and axicons at
  telecom wavelengths based on plasmonic metasurfaces}.
\newblock \emph{\bibinfo{journal}{Nano Lett.}} \textbf{\bibinfo{volume}{12}},
  \bibinfo{pages}{4932--4936} (\bibinfo{year}{2012}).

\bibitem{Arbabi2014arXiv}
\bibinfo{author}{Arbabi, A.} \& \bibinfo{author}{Faraon, A.}
\newblock \bibinfo{title}{Fundamental limits of ultrathin metasurfaces}.
\newblock \emph{\bibinfo{journal}{arXiv.org, e-Print Arch., Phys.
  arXiv:1411.2537}}  (\bibinfo{year}{2014}).

\bibitem{Aieta2015Science}
\bibinfo{author}{Aieta, F.}, \bibinfo{author}{Kats, M.~A.},
  \bibinfo{author}{Genevet, P.} \& \bibinfo{author}{Capasso, F.}
\newblock \bibinfo{title}{Multiwavelength achromatic metasurfaces by dispersive
  phase compensation}.
\newblock \emph{\bibinfo{journal}{Science}} \textbf{\bibinfo{volume}{347}},
  \bibinfo{pages}{1342--1345} (\bibinfo{year}{2015}).

\bibitem{Khorasaninejad2015NanoLett}
\bibinfo{author}{Khorasaninejad, M.} \emph{et~al.}
\newblock \bibinfo{title}{Achromatic metasurface lens at telecommunication
  wavelengths}.
\newblock \emph{\bibinfo{journal}{Nano Lett.}} \textbf{\bibinfo{volume}{15}},
  \bibinfo{pages}{5358--5362} (\bibinfo{year}{2015}).

\bibitem{Zhao2015SciRep}
\bibinfo{author}{Zhao, Z.} \emph{et~al.}
\newblock \bibinfo{title}{Multispectral optical metasurfaces enabled by
  achromatic phase transition}.
\newblock \emph{\bibinfo{journal}{Sci. Rep.}} \textbf{\bibinfo{volume}{5}},
  \bibinfo{pages}{15781} (\bibinfo{year}{2015}).

\bibitem{Cheng2015JOSAB}
\bibinfo{author}{Cheng, J.} \& \bibinfo{author}{Mosallaei, H.}
\newblock \bibinfo{title}{Truly achromatic optical metasurfaces: a filter
  circuit theory-based design}.
\newblock \emph{\bibinfo{journal}{J. Opt. Soc. Am. B}}
  \textbf{\bibinfo{volume}{32}}, \bibinfo{pages}{2115--2121}
  (\bibinfo{year}{2015}).

\bibitem{Oskooi2010CompPhys}
\bibinfo{author}{Oskooi, A.~F.} \emph{et~al.}
\newblock \bibinfo{title}{Meep: A flexible free-software package for
  electromagnetic simulations by the fdtd method}.
\newblock \emph{\bibinfo{journal}{Comput. Phys. Commun.}}
  \textbf{\bibinfo{volume}{181}}, \bibinfo{pages}{687--702}
  (\bibinfo{year}{2010}).

\bibitem{Liu2012CompPhys}
\bibinfo{author}{Liu, V.} \& \bibinfo{author}{Fan, S.}
\newblock \bibinfo{title}{S4 : A free electromagnetic solver for layered
  periodic structures}.
\newblock \emph{\bibinfo{journal}{Comput. Phys. Commun.}}
  \textbf{\bibinfo{volume}{183}}, \bibinfo{pages}{2233--2244}
  (\bibinfo{year}{2012}).

\end{thebibliography}

\clearpage

\newcommand{\beginsupplement}{%
        \setcounter{table}{0}
        \renewcommand{\thetable}{S\arabic{table}}%
        \setcounter{figure}{0}
        \renewcommand{\thefigure}{S\arabic{figure}}%
     }

\onecolumngrid

      \beginsupplement

\renewcommand{\figurename}{}
\renewcommand{\thefigure}{\textbf{Supplementary Information Figure S\arabic{figure}}}     

\section{Supplementary Information for ``Multiwavelength polarization insensitive lenses based on dielectric metasurfaces with meta-molecules"}

\section{S1. Discussion of deflection efficiency of blazed gratings designed with the proposed meta-molecule platform}
To understand the reasons behind the low efficiency of the lenses at 915 nm, two double wavelength blazed gratings were designed using the proposed meta-molecule scheme. One grating with a small deflection angle (5 degrees) and another one with a larger angle (20 degrees) were simulated at 915 nm using MEEP, and power loss channels were analyzed in both cases (Fig. S3). Both gratings were chosen to be 2 meta-molecules wide in the $y$ direction, so that periodic boundary conditions in this direction can be used in FDTD. The 5 degree grating is 322 lattice constants long in $x$ direction, while the 20 degree one is 146 lattice constants. The lengths are chosen such that the grating phases at 915 nm and 1550 nm are both almost repeated after the chosen lengths (Fig. S3a). An x-polarized plane-wave normally incident from the fused silica side was used as excitation in both simulations, and the transmitted and reflected electric and magnetic field intensities were calculated about a wavelength apart from the meta-molecules. The transmitted fields were further propagated using a plane wave expansion program, and the resulting fields in an area of length 30 $\mu$m around the center can be seen in Fig. S3b and S3c for 5 degree and 20 degree gratings, respectively. The field distributions outside of the areas shown here look similar to the ones shown. In both cases, a dominant plane wave propagating in the design direction is observed, along with some distortions. Angular distribution of power in transmission and reflection is analyzed using the Fourier transform of the fields. The resulting power distributions are shown in Fig. S3d and S3e for 5 degrees and 20 degrees, respectively. While the average power transmission of meta-molecules used in both gratings (found from the data in Fig. 2g) is slightly above 73$\%$, only 36$\%$ and 22$\%$ of the incident power is directed to 5 and 20 degrees for the corresponding gratings. The actual total transmitted powers are 56$\%$ and 50$\%$ for the 5 and 20 degree gratings, showing that an additional $\sim$20$\%$ of the power gets reflected as a result of the introduced aperiodicity. Because of the relatively large lattice constant, even a small aperiodicity can result in generation of propagating modes in the substrate, thus the reflection is considerably higher for the gratings than for a perfectly periodic lattice. From the 56$\%$ transmitted power in the 5 degree grating, 20$
\%$ is lost to diffraction to other angles. From Fig. S3b and S3c we can see there are distortions in the transmitted field. These distortions, mainly due to the low transmission amplitude of some of the meta-molecules and their phase errors result in the transmitted power being diffracted to other angles. Besides, it is seen that power loss to other angles both in reflection and transmission is higher for larger grating angles. This is due to the need for finer sampling of the wave front for waves with steeper angles. The lower efficiency for gratings with larger angles results in lower efficiency of lenses with higher numerical apertures which need bending light with larger angles.

\clearpage
\subsection{S.2 Supplementary Figures}
\begin{figure*}[htp]
\centering
\includegraphics{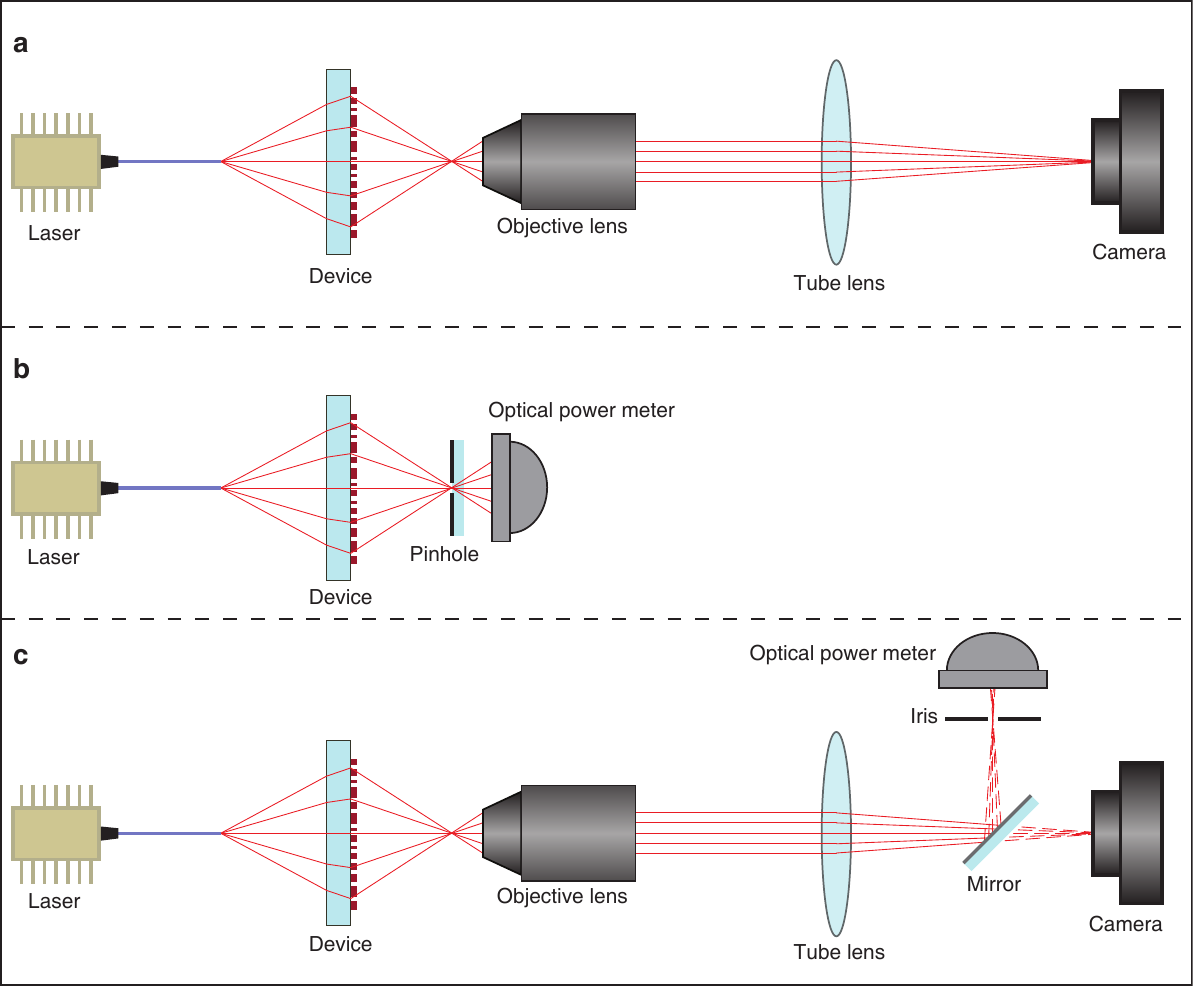}
\caption{\textbf{Measurement setups:} \textbf{a}, The measurement setup used to capture the focus pattern and the intensity distribution in different planes around focus. The laser source, fibers, tube lens, and camera were different in 915 nm and 1550 nm measurements. \textbf{b}, The measurement setup for measuring the efficiency of the lenses at 915 nm using a 20 $\mu$m pinhole in the focal plane. \textbf{c}, The setup for measuring focusing efficiency of the lens at 1550 nm using a 2 mm iris in the image plane of a $\sim$100X microscope.}
\end{figure*}

\begin{figure*}[htp]
\includegraphics{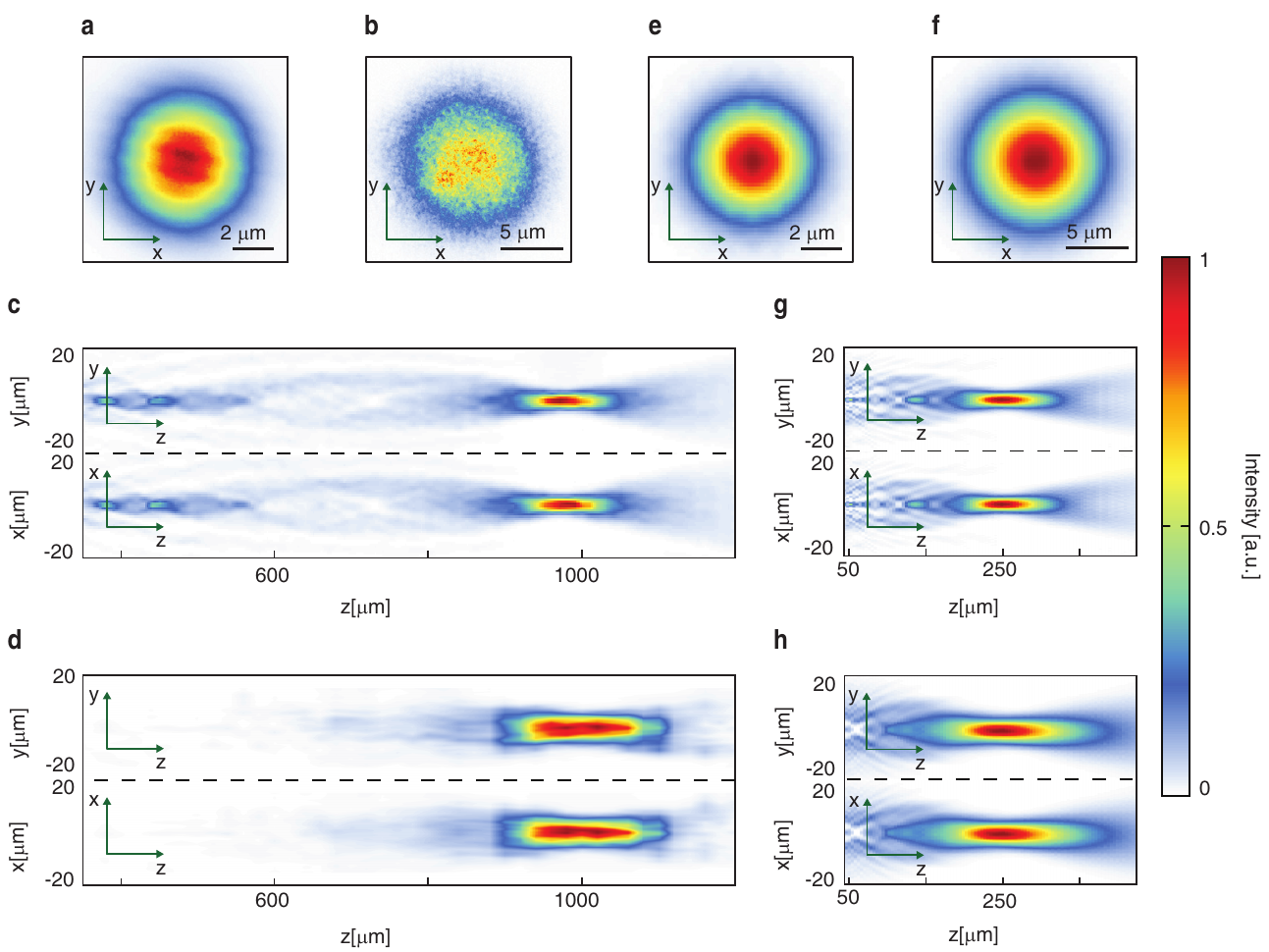}
\caption{\textbf{Measurement and simulation results for the lenses with a lower NA:} \textbf{a} and \textbf{b}, Measured intensity in the focal plane of a double wavelength lens (1000 $\mu$m focal length, 300 $\mu$m diameter) at 915 nm \textbf{a} and 1550 nm \textbf{b}. At 915 nm the lens actually focuses the light 980 $\mu$m away from its surface, so the focal spot shown here is imaged at $\approx$ 980 $\mu$m away from the surface. The error in focal distance is probably due to the approximation made in the mode diameter of the fiber (see Fig. S2), which affects the focusing distance of a low NA lens more than that of a high NA lens. \textbf{c}, Intensity measured in the axial planes of the lens for 915 nm. \textbf{d}, The same axial plots for 1550 nm. \textbf{e}, Simulated focal plane intensity of a lens with the same numerical aperture as the one shown in \textbf{a--d} but with a dimeter of 75 $\mu$m at wavelength of 915 nm, and \textbf{f} at 1550 nm. \textbf{g} and \textbf{h}, Simulated intensity profiles in the axial planes at 915 nm and 1550 nm, respectively, calculated for the same lens described in \textbf{e}.}
\end{figure*}

\begin{figure*}[htp]
\includegraphics{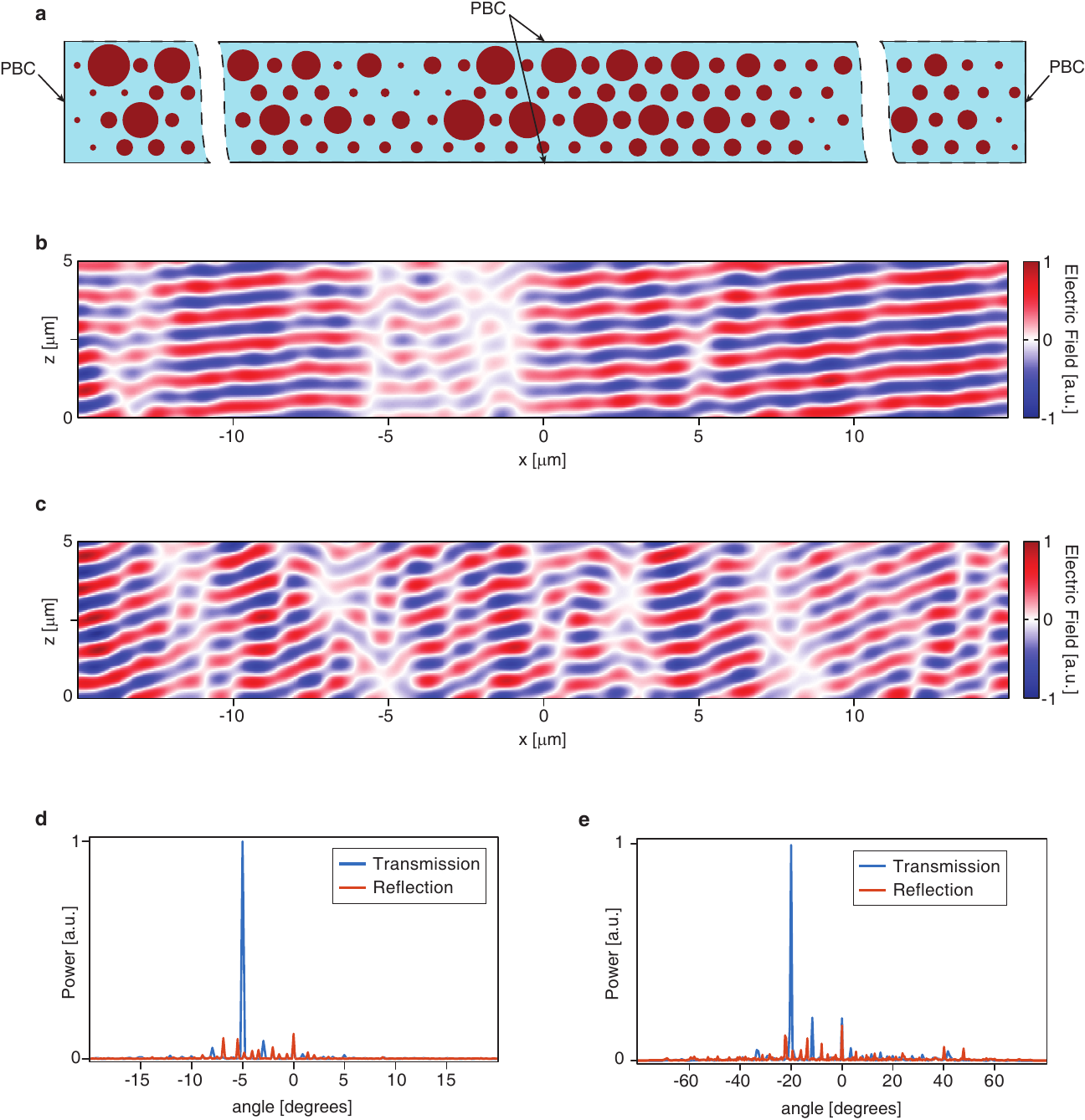}
\caption{\textbf{Double wavelengths blazed gratings based on the proposed meta-molecule design:} \textbf{a}, Schematic of the simulated grating. The 5 degree grating is 322 meta-molecules long, while the 20 degree one is 146  meta-molecules long. Periodic Boundary Conditions (PBC) was used in side boundaries and PML was used to terminate the simulation domain in top and bottom directions. \textbf{b} and \textbf{c}, Real part of the electric field a few wavelengths after the 5 degree \textbf{b} and 20 degree \textbf{c} gratings. \textbf{d} and \textbf{e}, Distribution of transmitted and reflected power in different angles for the 5 degree \textbf{d} and 20 degree \textbf{e} gratings.}
\end{figure*}

\end{document}